\documentclass{article}
\usepackage{graphicx} 
\usepackage{xcolor}
\usepackage{multirow}
\usepackage{booktabs}
\usepackage{tabularx}
\usepackage{hyperref}
\usepackage{subcaption}
\usepackage{adjustbox}
\usepackage{underscore}
\usepackage{orcidlink} 

\usepackage{amsmath}
\usepackage{bbm}
\usepackage{amsfonts}
\usepackage{bm}

\usepackage{tikz}
\usepackage{pgfplots}
\pgfplotsset{compat=1.18}
\NewDocumentCommand{\citep}{O{} O{} m}{%
  \parencite[#1][#2]{#3}%
}

\NewDocumentCommand{\citet}{O{} O{} m}{%
  \textcite[#1][#2]{#3}%
}
\usetikzlibrary{
shapes,
arrows, shapes.geometric,
shapes.arrows,
positioning,
decorations.pathreplacing,
angles,quotes,
fit,
calc, 
shapes.symbols} 
\NewDocumentCommand\Cycle{O{} m m m O{} m m}{%
  \draw[#1,<->, very thick, >=stealth'](#2.{#3+asin(#6/(#4*1.41))}) arc (180+#3-45:180+#3-45-#7:#6/2) #5;
}

\usepackage[ruled,vlined,linesnumbered,longend]{algorithm2e}
\makeatletter
\renewcommand{\algocf@captiontext}[2]{\AlCapNameSty{\AlCapNameFnt{}#2\endgraf}}%
\makeatother
\SetKwFunction{GenerateData}{generate\_data}
\SetKwFunction{EstimateModel}{estimate\_model}
\SetKwFunction{EvaluatePredictions}{evaluate\_predictions}
\SetKwFunction{isValid}{is\_valid}

\def\eg{e.\,g.,}
\def\ie{i.\,e.,}

\usepackage[abbreviations,nolangwarn]{glossaries-extra}
\setabbreviationstyle{long-short}
\newabbreviation{sem}{SEM}{structural equation modeling}
\newabbreviation[shortplural={DGPs}]{dgp}{DGP}{data generating process}
\newabbreviation[first={partial least squares path modeling \citep[PLS-PM,][]{Wold1975, Wold1982,Tenenhaus2005PLSModeling}}
]{plspm}{PLS-PM}{partial least squares path modeling}
\newabbreviation[first={maximum likelihood \citep[ML,][]{Joreskog1969AAnalysis}}
]{ml}{ML}{maximum likelihood}
\newabbreviation{iqr}{IQR}{inter-quartile range}
\newabbreviation[shortplural={RMSEs}]{rmse}{RMSE}{root mean squared error}
\newabbreviation{mse}{MSE}{mean squared error}
\newabbreviation{mae}{MAE}{mean absolute error}
\newabbreviation{pls}{PLS}{partial least squares}
\newabbreviation[first={consistent partial least squares \citep[PLSc,][]{Dijkstra2015ConsistentModeling}}
]{plsc}{PLSc}{consistent partial least squares}
\glsunsetall
\usepackage[
backend=biber,
style=apa,
useprefix=true,
date=year,
url=false,
isbn=false
]{biblatex}
\DeclareSourcemap{
  \maps[datatype=bibtex]{
    \map{
      \step[fieldset=issn, null]
    }
  }
}

\title{
Out-of-sample prediction in SEM:\\ A simulation-based comparison of maximum likelihood and (consistent) partial least squares
}
\author{Jonas Bauer\orcidlink{0000-0003-1759-7889} \and Florian Schuberth\orcidlink{0000-0002-2110-9086} \and Christiane Fuchs\orcidlink{0000-0003-3565-8315} \and Tamara Schamberger\orcidlink{0000-0002-7845-784X}}

\begin{document}
\date{}
\maketitle
\begin{abstract}
Structural equation modeling (SEM) is widely used to study relationships among unobserved constructs.
Over recent years, the interest in computing predictions with SEM has grown.
Different prediction procedures have been proposed for this purpose, including MLpredict, which is based on maximum likelihood (ML) estimation, and PLS(c)predict, which is based on partial least squares (PLS).
The out-of-sample predictive performances of these procedures have not yet been compared. 
In our work, we conduct a comprehensive Monte Carlo study to compare out-of-sample predictions of MLpredict and PLS(c)predict for indicators associated with endogenous constructs, so-called operative predictions.
Here, we vary key design factors regarding, for instance, the strengths of structural relations, the training sample size, and the construct types in the data generating processes, which determine the presence or absence of construct misspecification. 
Across the investigated conditions, MLpredict and PLS(c)predict exhibit similar predictive performance according to two assessment criteria, indicating that neither procedure is uniformly superior for operative prediction. 
Out-of-sample predictive performance is driven mainly by the strength of the relations between the variables, whereas training sample size and construct misspecification have comparatively little influence on it.
An exception is the occasionally inflated prediction error of PLScpredict, which is in line with previous findings for ordinal indicators.
Overall, our results suggest that, in the studied setting, fine-grained methodological differences between traditional SEM and PLS-based procedures matter less for out-of-sample predictive performance than properties of the data generating process, such as the coefficient of determination of endogenous constructs. 
\end{abstract}

\section{Introduction}
Across disciplines, researchers are interested in unobserved variables \citep[\eg][]{DeCarvalho2014ApplicationsResearchMan, Gao2023TheTransformation}.
An established statistical method for analyzing such variables is structural equation modeling \citep[SEM,][]{Bollen1989}.
The roots of \gls{sem} date back more than a century \citep{Pearson1901LIII.Space, Wright1918OnFactors, Tarka2018AnSciences}, and continuous methodological refinements have made it applicable to a wide range of multivariate statistical problems, such as cross-sectional models \citep{Bollen1989}, longitudinal models \citep{Bollen2005LatentPerspective}, fixed and random effects models \citep{Mayer2012AComponents, Muthen2002LatentMplus}, and models with different types of constructs \citep{Schamberger2025Factor-Framework}.

Traditionally, \gls{sem} is used for explanatory modeling, \ie\ theory-based model specification and empirical hypothesis testing \citep[\eg][]{Hayduk2007TestingModels}.
Accordingly, the assessment of 
model fit plays a central role in this type of \gls{sem} \citep{ Schumacker2012AModeling}, with numerous advancements such as novel fit measures \citep{Foldnes2025ImprovedModels}, performance evaluation thresholds \citep{Marsh2013GoodnessModels}, and approaches for confirmatory composite analysis \citep[\eg][]{Schuberth2018ConfirmatoryAnalysis, Schamberger2023b}.
Next to explanatory modeling, \gls{sem} can also be used for predictive modeling \citep{Shmueli2016TheModels}.
That is, \gls{sem} is used to find the most plausible values of the target variables for new (out-of-sample) observations, given the observed values of their predictors \citep{Shmueli2010ToPredict}.
Compared to explanatory SEM, predictive SEM is a more recent research area, characterized by a comparatively smaller, yet steadily developing, scientific literature.

In predictive \gls{sem}, different procedures have been developed to compute out-of-sample predictions. 
For instance, \emph{PLSpredict} \citep{Shmueli2016TheModels} has been introduced to generate predictions from models estimated by \glsfirst{plspm}.
However, \gls{plspm} is known to produce inconsistent parameter estimates if constructs are modeled as common factors \citep{Dijkstra1983SomeMethods}.
To address this issue, \glsfirst{plsc} was introduced, which corrects for the attenuation bias in the \gls{plspm} estimates and produces consistent estimates for common factor models\footnote{Although \gls{plsc} has been extended to estimate common factors with random measurement errors that might be correlated within a block of indicators \citep{Rademaker2019MeasurementSquares}, it is limited in terms of model specification compared to \gls{ml} \citep{Schuberth2023Partial2021}.}. 
In line with the literature, we refer to prediction based on \gls
{plsc} as \emph{PLScpredict} \citep{Schamberger2023aMan} and refer to PLS-based prediction in general as PLS(c)predict. 
In addition, a prediction procedure for out-of-sample prediction based on traditional \gls{sem} estimators, \eg\ using the \glsfirst{ml} estimator, has been developed \citep{deRooij2023SEM-BasedPredictionsMan}.
This procedure uses the predictive distribution of the target variables, which is implied by the model under the assumption of normality of the indicators.
In analogy to PLSpredict, we refer to this procedure as \emph{MLpredict}.

Despite the emergence of these different prediction procedures, their performance has not yet been systematically compared.
Especially in situations with construct misspecification, \ie\ where the specified constructs in the model do not match those in the \glsfirst{dgp}, estimators based on \gls{plspm} and \gls{ml} yield a poor explanatory performance, \ie\ inconsistent and biased parameter estimates \citep[\eg][]{Sarstedt2016, Rhemtulla2020,Bauer2025MisspecificationsCompositesArxiv}.
In contrast, simulation studies on predictive \gls{sem} have largely assessed the adequacy of individual procedures \citep[\eg][]{Evermann2016AssessingEstimators,deRooij2023SEM-BasedPredictionsMan} and consequently do not offer a systematic comparison across them.
To date, the out-of-sample predictive performance of MLpredict has not yet been systematically contrasted with other established procedures.
Such rigorous comparisons are especially important because methods well suited for explanation are not necessarily equally capable of accurate prediction \citep{DelGiudice2024TheLearning}.

This article addresses this research gap by comparing MLpredict to \linebreak PLS(c)predict with respect to computing out-of-sample prediction of indicators associated with endogenous constructs in \gls{sem}. Our Monte Carlo study provides novel insights into the performance of these procedures.
We also examine whether previously found severe consequences of construct misspecification on the explanatory performance extend to poor predictive performance.

The following section describes the different procedures for computing out-of-sample predictions in \gls{sem} in more detail.
To compare these procedures, we conduct a Monte Carlo study and summarize its key results in Section~\ref{sec:simulation}.
The potential causes for these findings and their implications are discussed in Section \ref{sec:discussion}.
Finally, Section \ref{sec:conclusion} concludes this article.

\section{Out-of-sample prediction in SEM} \label{sec:prediction}

Out-of-sample prediction aims to predict certain target variables based on predictor variables for new observations \citep{Shmueli2010ToPredict}.
To formally differentiate the different prediction procedures, we introduce the following notation: 
Suppose that the model parameters~$\bm\theta$ have been previously estimated and predictions shall be made for \emph{one} new individual, which is characterized by a random vector of indicators~$\bm{\tilde x}$.
This random vector consists of random variables~$\bm{\tilde x}^{\text{ex}}$ and $\bm{\tilde x}^{\text{en}}$, which denote  the indicators associated with the exogenous constructs~$\bm\eta^{\text{ex}}$ and the endogenous constructs~$\bm\eta^{\text{en}}$, respectively, such that $\bm{\tilde x} = \left(\bm{\tilde x}^{\text{ex}},\bm{\tilde x}^{\text{en}}\right)^\top$.
Exogenous constructs do not have antecedents in the structural model, whereas endogenous constructs are explained by at least one other variable in the structural model.
The prediction of indicator values $\bm{\tilde x}^{\text{en}}$ associated with endogenous constructs using $\bm{\tilde x}^{\text{ex}}$ is called \emph{operative prediction} \citep[][]{Shmueli2016TheModels}.
This article focuses on operative prediction for two reasons: First, MLpredict has been developed and implemented only for this prediction type.
Second, only this prediction type can be evaluated in applications since both predictor and target variables are indicators and thus observed.

The proposed procedures to perform out-of-sample prediction in \gls{sem} belong to two distinct families: MLpredict and PLS(c)predict. 
MLpredict is fully probabilistic, but reliant on assumptions regarding the model structures and probability density functions, whereas PLS(c)predict is based on a deterministic algorithm. 
For both procedures, the model parameters are estimated based on a training sample $\bm X$ in the first step, resulting in the vector of parameter estimates~$\hat{\bm\theta}$. 
Subsequently, a prediction of MLpredict is the conditional mean of the target variables derived by imposing distributional assumptions on the indicators, \ie\ using properties of generative models. 
MLpredict has been developed specifically for operative prediction using traditional \gls{sem} estimators that minimize the discrepancy between the indicators' empirical and model-implied variance-covariance matrices \citep{deRooij2023SEM-BasedPredictionsMan}.
In principle, however, it can be extended to other prediction types and combined with any estimator. 
Based on the estimated parameters, MLpredict then assumes that a new observation~$\bm{\tilde x}$ is multivariate normally distributed with model-implied mean~$\bm\mu(\hat{\bm \theta})$ and variance–covariance matrix~$\bm\Sigma(\hat{\bm \theta})$.
For the sake of clarity, we drop the dependencies on $\hat{\bm\theta}$ hereafter, and denote model-implied characteristics by a hat (\eg\ $\hat{\bm\mu}=\bm\mu(\hat{\bm \theta})$).
Moreover, the estimated variance-covariance matrix~$\bm{\hat\Sigma}$ can be partitioned as follows:
\[
  \bm{\hat\Sigma}
  =
  \begin{pmatrix}
    \bm{\hat\Sigma}_{\bm{\tilde x}^{\text{ex}}, \bm{\tilde x}^{\text{ex}}}
    & \bm{\hat\Sigma}_{\bm{\tilde x}^{\text{ex}}, \bm{\tilde x}^{\text{en}}} \\
    \bm{\hat\Sigma}_{\bm{\tilde x}^{\text{en}}, \bm{\tilde x}^{\text{ex}}}
    & \bm{\hat\Sigma}_{\bm{\tilde x}^{\text{en}}, \bm{\tilde x}^{\text{en}}}
  \end{pmatrix}.
\]
MLpredict subsequently predicts the indicators~$\bm{\tilde x}^{\text{en}}$ based on the available information in $\bm{\tilde x}^{\text{ex}}$ via the conditional distribution of~$\bm{\tilde x}^{\text{en}}$ given~$\bm{\tilde x}^{\text{ex}}$.
More precisely, if the joint distribution of the indicators is normal, then this conditional distribution is also multivariate normal.
The conditional mean~$\bm{\hat\mu}_{\bm{\tilde x}^{\text{en}} \mid \bm{\tilde x}^{\text{ex}}}$ of this distribution is the expected value of the target variable and given as
\begin{equation}
\label{eq:ml-cond-mean}
  \bm{\hat\mu}_{\bm{\tilde x}^{\text{en}} \mid \bm{\tilde x}^{\text{ex}}}
  =
  \bm{\hat\mu}_{\bm{\tilde x}^{\text{en}}}
  +
  \bm{\hat\Sigma}_{\bm{\tilde x}^{\text{ex}}, \bm{\tilde x}^{\text{en}}}^\top
  \bm{\hat\Sigma}_{\bm{\tilde x}^{\text{ex}}, \bm{\tilde x}^{\text{ex}}}^{-1}
  (\bm{\tilde x}^{\text{ex}} - \bm{\hat\mu}_{\bm{\tilde x}^{\text{ex}}}).
\end{equation}
The conditional mean in Equation~\eqref{eq:ml-cond-mean} can thus be decomposed into its unconditional counterpart~$\bm{\hat\mu}_{\bm{\tilde x}^{\text{en}}}$, a structural relationship term~$\bm{\hat\Sigma}_{\bm{\tilde x}^{\text{ex}}, \bm{\tilde x}^{\text{en}}}^\top \bm{\hat\Sigma}_{\bm{\tilde x}^{\text{ex}}, \bm{\tilde x}^{\text{ex}}}^{-1}$ that encodes the dependencies between the indicators~$\bm{\tilde x}^{\text{ex}}$ and those to be predicted, and lastly the new information~$\bm{\tilde x}^{\text{ex}} - \bm{\hat\mu}_{\bm{\tilde x}^{\text{ex}}}$, which reflects how the observed entries in $\bm{\tilde x}^{\text{ex}}$ deviate from their expectation.
Point predictions for~$\bm{\tilde x}^{\text{en}}$ are then given by this conditional mean~$\bm{\hat\mu}_{\bm{\tilde x}^{\text{en}} \mid \bm{\tilde x}^{\text{ex}}}$.
Moreover, the prediction uncertainty can be quantified using $\bm{\hat\Sigma}_{\bm{\tilde x}^{\text{en}} \mid \bm{\tilde x}^{\text{ex}}}$.
For further information on MLpredict, we refer to \citet{deRooij2023SEM-BasedPredictionsMan}.

Another procedure for performing out-of-sample prediction in \gls{sem} is PLSpredict \citep{Shmueli2016TheModels}. 
PLSpredict is a deterministic algorithm that does not rely on explicit distributional assumptions for the indicators.
Instead, it uses estimates of weights~$\hat{\bm W}$, factor loadings~$\hat{\bm \Lambda}$, and path coefficients~$\hat{\bm B}$ obtained by \gls{plspm} based on the training sample.
These estimates are derived iteratively in \gls{plspm} by alternating between building linear combinations of indicators as proxies for the constructs, \ie\ construct scores, and then estimating model parameters based on these scores \citep[\eg][]{Tenenhaus2005PLSModeling}.
The first step to compute predictions with PLSpredict is standardization because the parameter estimates derived from \gls{pls} are based on standardized variables. 
That is, the $j$-th entry of the new individual~$\bm{\tilde x}^{\text{ex}}$ from the test sample is standardized using the empirical mean~$\bar{x}^{\text{ex}}_j$ and the empirical variance~$S^{\text{ex}}_{jj}$ of the training sample, yielding the standardized indicators~$\bm{\tilde z}^{\text{ex}}, \tilde{z}^{\text{ex}}_j=  \left(\tilde x^{\text{ex}}_j - \bar{x}^{\text{ex}}_j\right)/\sqrt{S^{\text{ex}}_{jj}}$.
Consequently, $\bm{\tilde z}^{\text{ex}}$ are the standardized predictors and~$\bm{\tilde z}^{\text{en}}$ are the standardized target variables to be predicted.
Scores for the exogenous constructs are then given by the linear combination
$$\hat{\bm{\eta}}^{\text{ex}} = (\hat{\bm W}^{\text{ex}})^{\top} \bm{\tilde z}^{\text{ex}},$$ 
where $\hat{\bm W}^{\text{ex}}$ consists only of the rows concerning the indicators associated with $\bm{\eta}^{\text{ex}}$.
Using the estimated path coefficients~$\hat{\bm B}$, the construct scores are passed along the directed paths of the structural model to obtain the score of the endogenous constructs~$\bm{\hat\eta}^{\text{en}}$.
The standardized predicted values~$\hat{\bm{\tilde z}}^{\text{en}}$ are then obtained using the estimated factor or composite loadings and construct scores, via
$$\hat{\bm{\tilde z}}^{\text{en}} = \bm{\hat\Lambda}^{\text{en}} \bm{\hat\eta}^{\text{en}},$$
where $\bm{\hat\Lambda}^{\text{en}}$ consists only of the columns concerning the endogenous constructs~$\bm{\hat\eta}^{\text{en}}$.
The final point prediction~$\hat{\bm{\tilde x}}^{\text{en}}$ is obtained by bringing $\hat{\bm{\tilde z}}^{\text{en}}$ to the original scale via~$\hat{\tilde{x}}^{\text{en}}_j =  \hat{\tilde{z}}^{\text{en}}_j \sqrt{S^{\text{en}}_{jj}} +
\bar{x}^{\text{en}}_j$, where $\bar{x}^{\text{en}}_j$ and $\sqrt{S^{\text{en}}_{jj}}$ are again calculated from the training sample.
More details on PLSpredict can be found in \citet{Shmueli2016TheModels}.

PLSpredict is based on \gls{plspm} estimates, irrespective of the operationalizations of constructs in the model. 
When constructs are modeled as common factors, however, the PLS-PM estimates of the loadings~$\hat{\bm \Lambda}$ and path coefficients~$\hat{\bm B}$ are known to be inconsistent \citep{Rigdon2016, hui1982consistency, dijkstra1981latent}.
To remedy this, \gls{plsc} has been introduced, corrects for attenuation bias in these parameters.
When predictions are based on PLSc estimates instead of \gls{plspm}, we refer to the procedure as \emph{PLScpredict} \citep{Schamberger2023aMan}.

The extent to which \gls{plsc}'s parameter estimation consistency, relative to \gls{plspm}, translates into differences in predictive performance received little attention in the \gls{sem} literature. 
For scenarios with ordinal indicators, the disattenuation used in PLScpredict did not improve the predictive performance \citep{Schamberger2023aMan}.
We therefore investigate the effects for continuous indicators in our simulation. 
More generally, we compare the two families of procedures in terms of predictive performance in a comprehensive Monte Carlo study in order to provide practical guidance for applied scientists.

\section{Monte Carlo study} \label{sec:simulation}

To evaluate and compare the predictive performance of MLpredict and \linebreak PLS(c)predict, we conduct a simulation study.
Our primary focus lies on two research questions: (1) To what extent do the procedures differ in their operative prediction performance, and (2) what are the implications of construct misspecification on the predictive performance, where misspecification means a mismatch between the construct types in the specified model and the \gls{dgp}?

\subsection{Study design} \label{sec:design}
In our simulation study, we generate data from structural equation models with four constructs.
Specifically, the two constructs~$\eta_1$ and $\eta_2$ are structurally exogenous, whereas the remaining two constructs~$\eta_3$ and $\eta_4$ are endogenous.
Each construct is associated with $K=3$ indicators.
Throughout our study, we set population parameters such that all four constructs have a variance of one.
Their structural relations are defined as follows:
\begin{align*}
    \eta_3 & = \beta_{1} \eta_1 + \beta_{2} \eta_2 + \zeta_3 \\
    \eta_4 & = \beta_{3}  \eta_2 + \beta_{4}  \eta_3 + \zeta_4.
\end{align*}

To evaluate the predictive performance of MLpredict and PLS(c)predict in more detail, we vary several design conditions.
Specifically, we vary the strength of the structural relations, \ie\ the correlation~$\rho$ between the two exogenous constructs~$\eta_1$ and $\eta_2$, and the path coefficients~$\beta_j$.
We set~$\rho$ to 0.2, 0.3, or 0.4, with larger values indicating greater resemblance between the exogenous constructs.
In addition, we vary the coefficients of determination~$R^2$ for the two endogenous constructs~$\eta_3$ and $\eta_4$.
The $R^2$ characterizes how much of the dependent constructs' variance is explained by the antecedent constructs, \ie\ how strongly these constructs are related.
We use the same $R^2$ value for both constructs in each simulation condition.
Because both endogenous constructs have two antecedents, different combinations of path coefficients can produce the same $R^2$.
We therefore consider both equal and unequal relative importance of the antecedents in our model. $\beta_{1}$ and $\beta_{2}$ are equal across all simulation conditions so that $\eta_1$ and $\eta_2$ contribute equally to $\eta_3$, whereas $\beta_{3}$ and $\beta_{4}$ are unequal across all simulation conditions so that $\eta_2$ and $\eta_3$ affect $\eta_4$ differently.
A derivation of the exact formulas for path coefficients~$\beta_j$ and the variances of the disturbance terms~$\zeta_3$ and $\zeta_4$ from the coefficients of determination is provided in Appendix \ref{appendix:path-coefs}.
In our study, we set the $R^2$ values to 0.2, 0.3, or 0.4.
The resulting population path coefficients are displayed in Table~\ref{tab:populationParametersStructural}.
\begin{table}[htp!]
    \centering
    \begin{tabular}{l|lll|lll|lll}
    \toprule
    $\rho$& \multicolumn{3}{c}{0.2} & \multicolumn{3}{c}{0.3} & \multicolumn{3}{c}{0.4}\\
    $R^2$& 0.2 & 0.3 & 0.4 & 0.2 & 0.3 & 0.4 & 0.2 & 0.3 & 0.4\\
    \midrule
    $\beta_{1},\beta_{2}$&0.289 & 0.354 & 0.408 & 0.277 & 0.340 & 0.392 & 0.267 & 0.327 & 0.378\\
    $\beta_{3}$& 0.318 & 0.372 & 0.414 & 0.315 & 0.369 & 0.410 & 0.313 & 0.365 & 0.405 \\
    $\beta_{4}$& 0.224 & 0.274 & 0.316 & 0.224 & 0.274 & 0.316 & 0.224 & 0.274 & 0.316 \\
   \bottomrule
    \end{tabular}
    \caption{Population path coefficients for different correlations between the exogenous constructs and different $R^2$ values for the endogenous constructs.}
    \label{tab:populationParametersStructural}
\end{table}

Besides the strengths of the structural relations, we further investigate whether the type of the constructs affects the predictive performance. 
We refer to this design factor as \emph{configuration} and consider the following three configurations  for $\eta_1$, $\eta_2$, $\eta_3$ and $\eta_4$ in the \glspl{dgp}:
(1)\ all constructs are common factors (FFFF), (2)\ all constructs are composites (CCCC) similar to those in (1), or (3)\ all constructs are composites (CCCC) dissimilar to those in (1).
In case of common factors, the relations between the construct and its indicators are given as follows: 
\begin{align*}
    x_{j,1} &= \lambda_{j,1} \eta_j + \varepsilon_{j,1}\\
    x_{j,2} &= \lambda_{j,2} \eta_j + \varepsilon_{j,2} \\
    x_{j,3} &= \lambda_{j,3} \eta_j + \varepsilon_{j,3}
\end{align*}
Each factor loading~$\lambda_{j,k}$ captures the influence of the common factor $\eta_j$ on the respective indicator~$x_{j,k}$.
The factor loadings are set to $\bm\lambda_j =(0.9, 0.8, 0.7)^\top$. 
The variance in an indicator that cannot be explained by the common factor is captured in the corresponding measurement error $\varepsilon_{j,k}$.
In our study, these measurement errors are assumed to be uncorrelated with all other variables in the model and have a variance of $0.3$. 
Since the constructs in our study have a variance of one, all associated indicators are unstandardized and have decreasing correlations with the common factors.

For composites in our study, we distinguish two different cases: (i) those whose relationships with their indicators resemble the relationships implied by the common factor described above, and (ii) those whose relationships differ substantially, resulting in entirely different model-implied variances and covariances, including the variance-covariance matrix of the indicators, covariances between indicators and the construct, and the variance of the construct. 
If a construct is a composite, $\eta_j$ is a weighted linear combination of its $3$ indicators:
\begin{align*}
    \eta_j &= w_{1} x_{j,1}   + w_2 x_{j,2} + w_3  x_{j,3}
\end{align*}
The variances and covariances of the indicators forming a composite are unrestricted and denoted as $\sigma_{ij}$.
In the case of $\eta_j$ being a composite with high similarity to the respective common factor, the population parameters are set such that the correlations with its associated indicators~$\text{Cor}(x_{j,k}, \eta_j)$ are close to the correlations between the common factor and its associated indicators.
Moreover, (co-)variances between the indicators of that composite are set to be equal to those of the common factor indicators.
Lastly, the population parameters of similar composites are chosen such that each composite has unit variance.
These conditions can be fulfilled by setting the weights~$\bm w_j$ according to Equation \ref{app:compW} and the (co-)variances of the indicators according to Equation \ref{app:compSigma} in the Appendix.
In contrast, the parameters of dissimilar composites are chosen independently from the common factors.
Consequently, dissimilar composites have (1) a different model-implied variance-covariance matrix of their indicators, and (2) different covariances with their indicators, compared to the case of common factors.
This set-up leads to the population parameters given in Table~\ref{tab:populationParametersStructural}.

\begin{table}[ht]
\centering
\begin{tabular}{l r r}
\toprule
Type & \multicolumn{1}{c}{$\bm{w}_j$}  & \multicolumn{1}{c}{$\bm{\Sigma}_j$} \\
\midrule
Similar
&
$\displaystyle
\begin{pmatrix}
0.43 \\ 0.38 \\ 0.34
\end{pmatrix}$
&
$\displaystyle
\begin{pmatrix}
1.11 & 0.72 & 0.63 \\
0.72 & 0.94 & 0.56 \\
0.63 & 0.56 & 0.79
\end{pmatrix}$
\\[1em]
Dissimilar
&
$\displaystyle
\begin{pmatrix}
0.20 \\ 0.40 \\ 0.60
\end{pmatrix}$
&
$\displaystyle
\begin{pmatrix}
1.32 & 0.25 & 0.36 \\
0.25 & 1.22 & 0.47 \\
0.36 & 0.47 & 1.11
\end{pmatrix}$
\\
\bottomrule
\end{tabular}
\caption{Population parameters of CCCC \glspl{dgp} in our study that are similar (top) or dissimilar (bottom) to the FFFF \glspl{dgp}. Values are rounded to two decimal places. }
\end{table}

Additionally, we include the following design factors:
First, we vary the training sample size~$N\in \{100, 300, 500\}$ that is used for model estimation to account for sample effects on the predictive performance.
Second, we vary the distribution of the indicators to be either normal or non-normal.
While PLSpredict makes no distributional assumption on the variables, MLpredict is based on the assumption of normality of the indicators.
To assess the influence of a violation of this assumption, we include the case of non-normal indicator distributions with moderate skewness and kurtosis values \citep{Reinartz2009AnSEM}. 
That is, non-normal indicator values are created with a skewness of 1 and an excess kurtosis of 6, \ie\ a kurtosis of 9, using the method in \citet{Vale1983SimulatingDistributions}.
Overall, the simulation consists of $162$ design conditions\footnote{162 = 3 correlations between $\eta_1$ and $\eta_2$ $\times$ 3 $R^2$ for $\eta_3$ and $\eta_4$ $\times$
3 configurations of the \gls{dgp} $\times$
3 training sample sizes $\times$ 2 normality of data.}.

\begin{figure}[htp!]
    \centering
    \begin{minipage}{\linewidth}
        \input{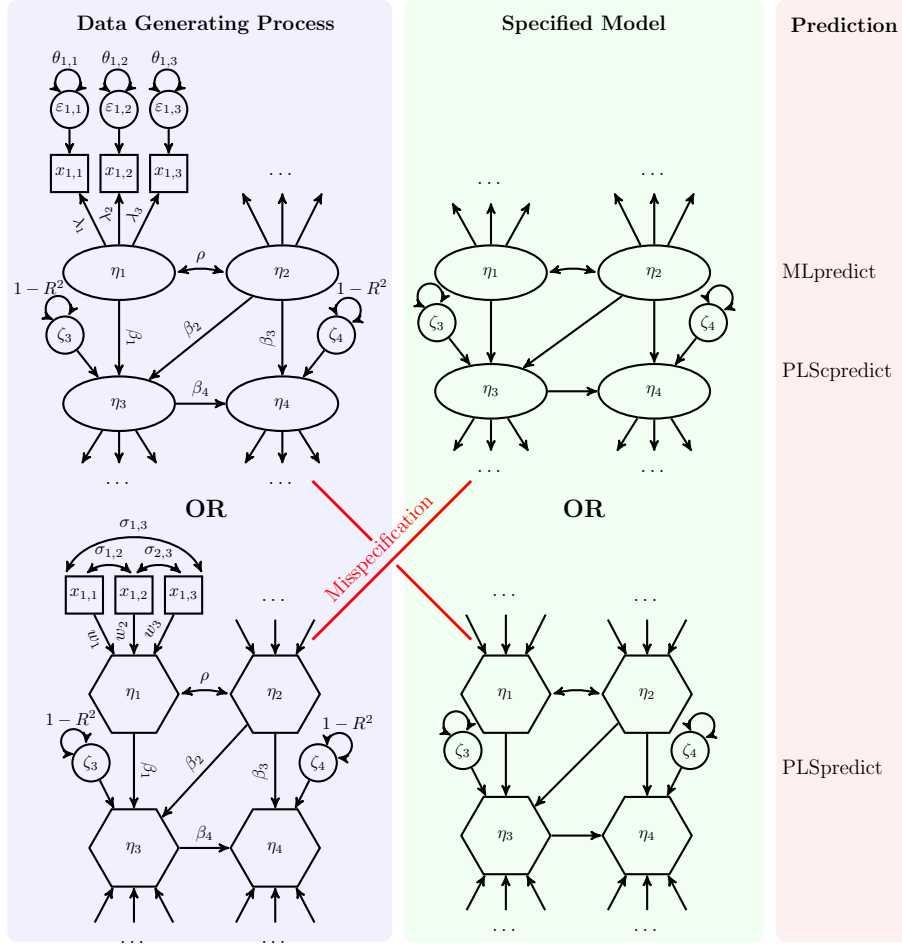}
    \end{minipage}
    \caption{
    Visualization of the combinations of \glspl{dgp} (column in blue), specified models (green), and prediction procedures (red) of our study. Common factors are represented as ovals, composites are represented as hexagons.}
     \label{fig:dgp:spec:pred}
\end{figure}

For all generated datasets, out-of-sample predictions are computed once using MLpredict, once using PLSpredict, and once using PLScpredict.
Each of these procedures comes with an assumption about the specified constructs.
Specifically, MLpredict has been introduced under the assumption that all constructs are specified as common factors.
In contrast, PLSpredict uses estimates obtained from \gls{plspm} and thus under the specification of all constructs as composites. 
Finally, PLScpredict can be used to compute out-of-sample predictions for models with both composites and common factors. 
In this study, however, the focus lies on a fair comparison with MLpredict and on the effect of the correction for disattenuation of \gls{plsc} on the predictive performance. 
Therefore, we apply PLScpredict only to models where all constructs are specified as common factors.
Figure~\ref{fig:dgp:spec:pred} depicts the conceptual levels of the \gls{dgp}, the specified model and the prediction procedure.
The red paths mark combinations where the construct types in the specified model and the \gls{dgp} differ, constituting a construct misspecification.
Analogously, the prediction procedures imply construct misspecification in our study when MLpredict and PLScpredict are used on CCCC \glspl{dgp}, or when PLSpredict is used on FFFF \glspl{dgp}. 
Consequently, our set-up enables us to evaluate the effects of construct misspecification on the predictive performance.

The steps of our Monte Carlo simulation are depicted in the pseudo-code Scheme~\ref{alg:mc-prediction} and are carried out in the statistical software \textsf{R} \citep[version \textsf{4.6.1};][]{RCoreTeam2026}.
For the sake of completeness and reproducibility, all \textsf{R} source files, the estimation results and Supplementary Material are available at \href{https://doi.org/10.5281/zenodo.23506099}{https://doi.org/10.5281/zenodo.23506099}.
In each such simulation run, we generate a training sample of the respective size and a test sample of additional 1,000 observations from the \gls{dgp} of the corresponding scenario using \texttt{simulateData()} from the package \texttt{lavaan} \citep[version \textsf{0.7-2};][]{lavaan2026, Rosseel2012}.
We then estimate parameters and compute predictions using MLpredict with the package \texttt{lavaan} \citep{lavaan2026,Rosseel2012} and PLSpredict and PLScpredict with the package \textsf{cSEM} \citep[version \textsf{0.6.1};][]{csem2020}.
To ensure that our prediction examination is based on proper solutions, we retain only admissible estimations, \ie\ runs in which the optimizer converged, all variance estimates were non-negative, and variance-covariance matrices of the constructs and the indicators (or measurement errors, respectively) were positive definite.
Only if all three estimations (with \gls{ml}, \gls{plspm} and \gls{plsc}) yield admissible estimates for the training sample, the Monte Carlo run is included for computing predictions and assessing their performance on the test data.
Specifically, each prediction procedure uses the indicators of the test data that are associated with exogenous constructs to predict the remaining indicators.
We iterate over these Monte Carlo runs until 500 admissible results are obtained for each simulation condition.

\begin{algorithm}[htbp!]
\KwData{Design matrix $D$}
\KwResult{Prediction results $R_{\rm ML}$, $R_{\rm PLS}$, and $R_{\rm PLSc}$ for all scenarios in $D$}
\DontPrintSemicolon

\Begin{
$M \gets 500$, $R_{\rm ML} \gets \emptyset$, $R_{\rm PLS} \gets \emptyset$, $R_{\rm PLSc} \gets \emptyset$\;
    \ForEach{scenario $s \in D$}{
        $admissible, attempts \gets 0$\;
        
        \While{$admissible < M$}{
            $attempts \gets attempts +1$\;
            $X_{\rm train} \gets \GenerateData(s)$\;
            \tcp{\texttt{Estimate parameters}}
            
            $\hat{\theta}_{\rm ML\ \ } \gets \EstimateModel(X_{\rm train}, \text{`ML'})$\;
            $\hat{\theta}_{\rm PLS\ } \gets \EstimateModel(X_{\rm train}, \text{`PLS', .disattenuate = F})$\;
            $\hat{\theta}_{\rm PLSc} \gets \EstimateModel(X_{\rm train}, \text{`PLSc', .disattenuate = T})$\;
            
            \eIf{$\isValid(\hat{\theta}_{\rm ML}) \And \isValid(\hat{\theta}_{\rm PLS})\And \isValid(\hat{\theta}_{\rm PLSc})$}{
            $X_{\rm test} \gets \GenerateData(s)$\;
            \tcp{\texttt{Predict new observations for indicators associated with endogenous constructs}}

            $pred_{\rm ML\ \ } \gets \EvaluatePredictions(X_{\rm test}, \hat{\theta}_{\rm ML})$\;
            $pred_{\rm PLS\ } \gets \EvaluatePredictions(X_{\rm test}, \hat{\theta}_{\rm PLS})$\;
            $pred_{\rm PLSc} \gets \EvaluatePredictions(X_{\rm test}, \hat{\theta}_{\rm PLSc})$\;
            $R_{\rm ML} \gets R_{\rm ML} \cup pred_{\rm ML}$\;
            $R_{\rm PLS} \gets R_{\rm PLS} \cup pred_{\rm PLS}$\;
            $R_{\rm PLSc} \gets R_{\rm PLSc} \cup pred_{\rm PLSc}$\;
                $admissible \gets admissible + 1$\;
                
            }{
                \tcp{retry; discard inadmissible fits}
            }
        }
        $D[s,attempts] \gets attempts$
    }
    \Return{$R_{\rm ML}$, $R_{\rm PLS}$, $R_{\rm PLSc}$, $D$}
}
\caption{Pseudo code of the Monte Carlo simulation on operative prediction of indicators associated with endogenous constructs (MLpredict, PLSpredict and PLScpredict).}
\label{alg:mc-prediction}
\end{algorithm}

To evaluate the predictive performance, we compare each procedure's predictions with the actual indicator values of the test sample from the respective run. 
From the variety of evaluation measures proposed to summarize such comparisons into a single value, we make use of two criteria that capture different aspects of the predictive performances.
First, we use the \glsfirst{rmse}\footnote{Note that this criterion is sometimes referred to as RMSEp in the context of prediction in the literature \citep[\eg][]{deRooij2023SEM-BasedPredictionsMan}.}.
This criterion is on the same scale as the data, which means it has no upper bound, and can be sensitive to outliers \citep{Hyndman2006AnotherAccuracy}. 
For that reason, we further use Theil's measure for prediction accuracy~$U_1$, which scales the \gls{rmse}. 
For the mean-centered indicators of our study, its denominator is the sum of the standard deviations of observed and predicted values. 
A $U_1$ of 0 denotes a perfect prediction, whereas 1 represents the same prediction, regardless of the predictor values \citep{Ahlburg1984ForecastDecomposition}.
Let $\bm x$ be an indicator to be predicted and $\hat{\bm x}$ the corresponding predictions, then the criteria for the $1,000$ observations in the test data are defined as follows:
\begin{align*}
\text{RMSE}(\bm x, \hat{\bm x}) &= \sqrt{\frac{1}{1,000} \sum_{i=1}^{1,000} \left(x_i -\hat x _i\right)^2} \\ 
U_1(\bm x, \hat{\bm x}) & = \frac{\text{RMSE}(\bm x, \hat{\bm x})}{\sqrt{\frac{1}{1,000}\sum_{i=1}^{1,000} x_i^2} +\sqrt{\frac{1}{1,000}\sum_{i=1}^{1,000} \hat x_i^2}} 
\end{align*}

\clearpage
\subsection{Results}
In this section, we report the results of our Monte Carlo study comparing the predictive performance of MLpredict, PLSpredict, and PLScpredict across the simulation conditions outlined above. 
We first examine the overall predictive performance for the considered \glspl{dgp} using the \gls{rmse}, and Theil’s $U_1$ criterion. 
We then investigate whether the performance varies systematically with the design factors.
Throughout this section, we pay particular attention to the consequences of construct misspecification on the performance of the three procedures.

\begin{figure}[ht!]
    \centering
    \includegraphics[width=\linewidth]{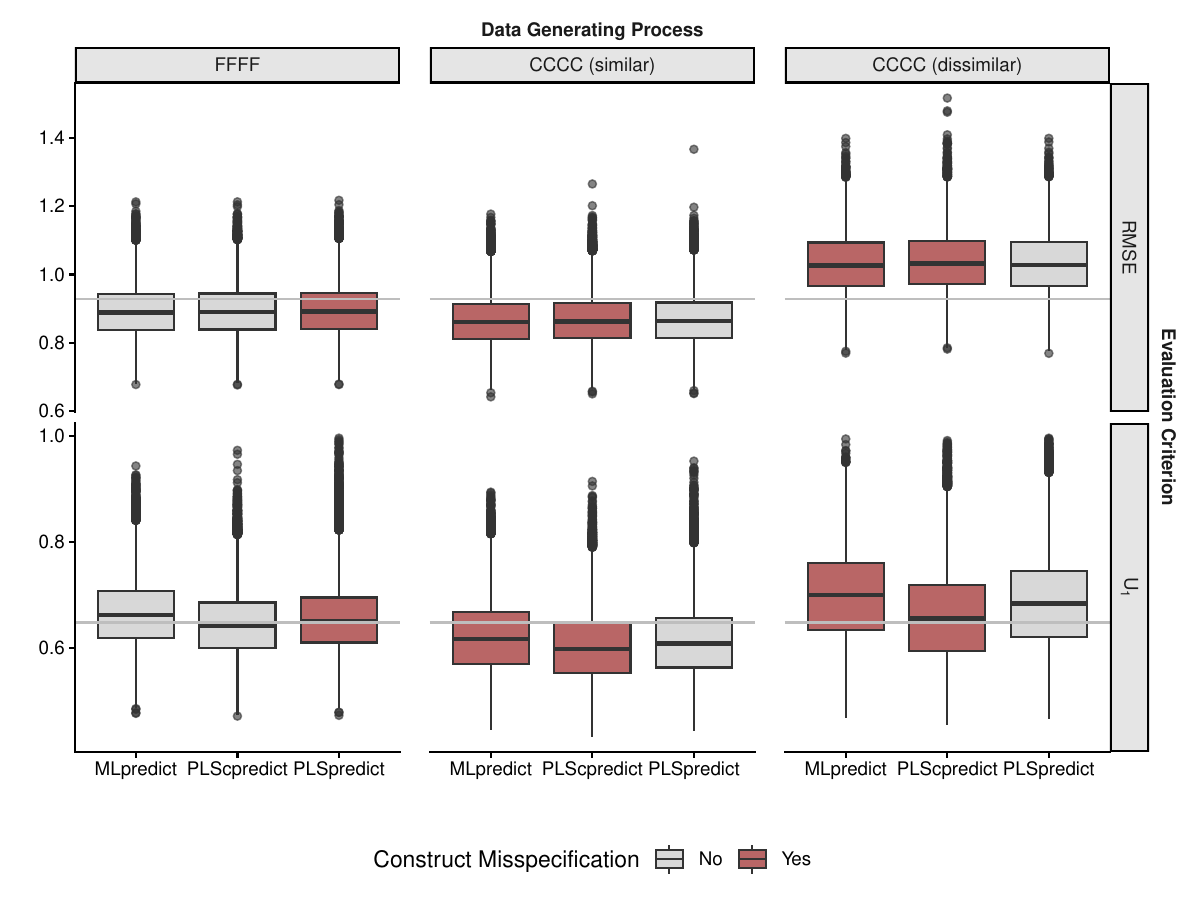}
    \caption{\gls{rmse}, and $U_1$ (in the rows) of MLpredict, PLSpredict, and PLScpredict (on the x axis) of all predicted indicators and all Monte Carlo conditions with data from common factor models (FFFF, left column), similar composite models (CCCC; center) and dissimilar composite models (CCCC; right). Red boxes indicate conditions with construct misspecification.}
    \label{fig:operative_all}
\end{figure}
The overall predictive performances of MLpredict, PLScpredict and PLSpredict are depicted in Figure \ref{fig:operative_all}. 
The \gls{dgp} configurations FFFF, similar CCCC, and dissimilar CCCC are distinguished in the columns and the two evaluation criteria are shown in the rows. 
Within each panel, the distributions of the performance measures for the three procedures are visualized as boxplots. 
Each box displays the \glsfirst{iqr}, containing the central 50\% of the performance measures.
The median value is highlighted by a horizontal line within each box.
The whiskers visualize the minimum and maximum values of the performance measure within 1.5 times the \gls{iqr}, while values beyond that range are treated as outliers and depicted as individual points.
The overall means for the performance criteria of our study are depicted as gray horizontal lines.
For each performance measure and each \gls{dgp} configuration, the three procedures exhibit generally a highly similar predictive performance:
For the \gls{rmse}, performances are almost identical.
Roughly 1\% of the Monte Carlo runs produce large prediction errors, as indicated by the outliers in the boxplots.
For PLScpredict, a small number of these outliers are particularly extreme in dissimilar CCCC \glspl{dgp}.
Under the $U_1$ criterion, PLScpredict yields slightly lower values than the other procedures.
Since all procedures have comparable \glspl{rmse}, the smaller $U_1$ indicates a larger prediction variance of PLScpredict.
Overall, however, the differences between the procedures remain small relative to the variation across simulation conditions. 
Certain combinations of prediction procedure and \glspl{dgp} constitute construct misspecification, as described above. 
Specifically, this is the case for the CCCC \glspl{dgp} using MLpredict and PLScpredict, and for the FFFF \glspl{dgp} using PLSpredict. 
These combinations are indicated by the red boxes in Figure~\ref{fig:operative_all}.
Comparing the red boxes with the gray boxes, where the construct type in the \gls{dgp} matches the construct type used for model estimation and prediction, reveals no systematic differences in the distributions of the performance measures. 
Thus, the overall predictive performance in our study is largely unaffected by whether the constructs are correctly specified or misspecified.
However, when the composites are similar to the common factor case (center column), the predictive performance is also similar to that of FFFF \glspl{dgp}.
In contrast, the dissimilar CCCC \glspl{dgp} yield larger prediction errors and a larger variability in the evaluation criteria.

\begin{figure}[htp!]
    \centering
    \includegraphics[width=\linewidth]{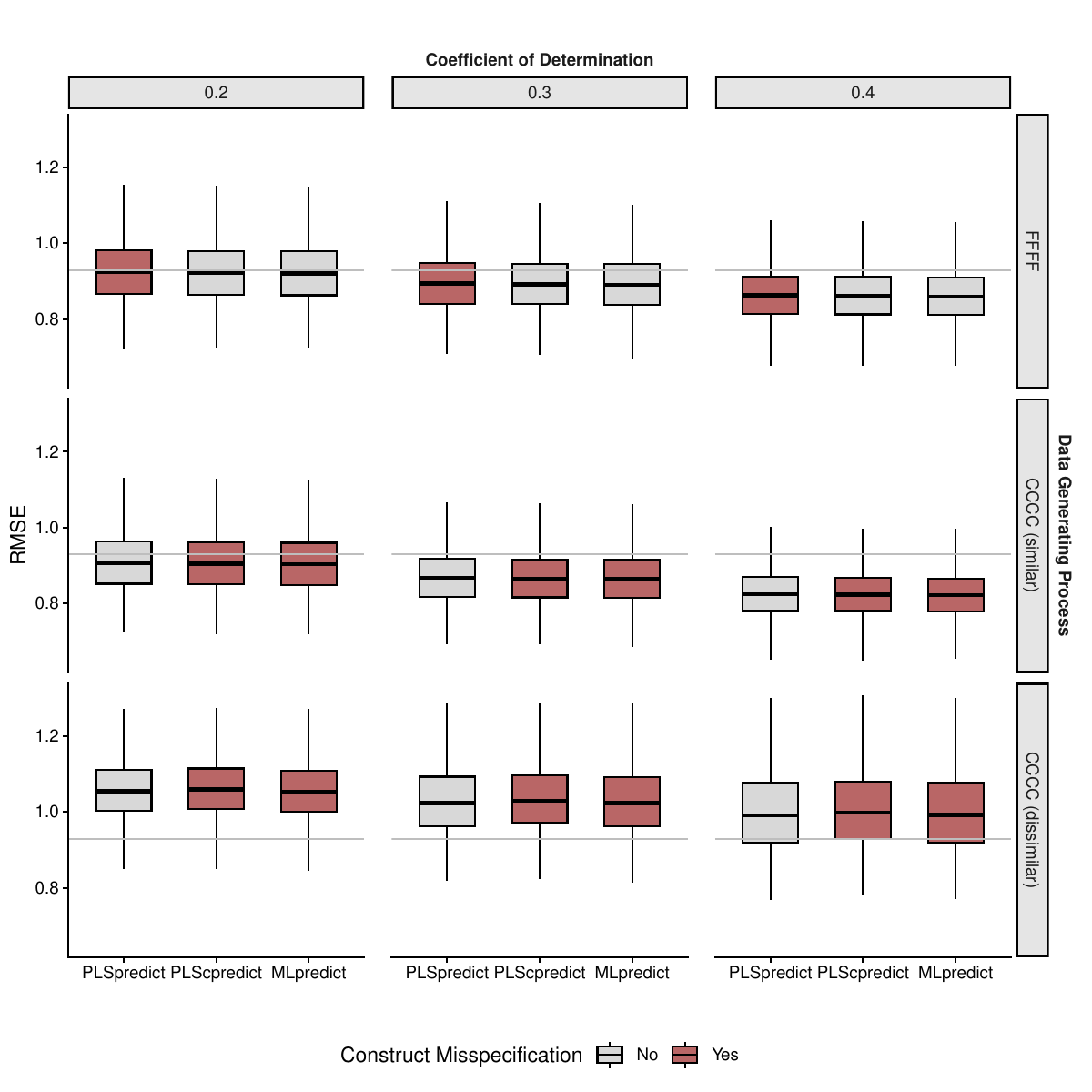}
    \caption{\gls{rmse} of operative predictions for all indicators associated with endogenous constructs with respect to the $R^2$ values (in the columns) in our study. Red boxes indicate conditions involving construct misspecification. The gray horizontal line represents the overall mean \gls{rmse} value of our study.}
    \label{fig:r2}
\end{figure}
Next, we assess the effects of the design factors on the predictive performance.
Because the results for the performance measures differ only marginally, we focus hereafter on the \gls{rmse} \citep[\eg\ as done in][]{deRooij2023SEM-BasedPredictionsMan}.
Moreover, we focus on the most influential design factor, namely the coefficients of determination of endogenous constructs.
All remaining criteria, \ie\ the size of the training sample, the correlation among the exogenous constructs, and the normality of the data, have little effect on the predictive performance, we present their results in Appendix \ref{app:sim_res}.
Figure \ref{fig:r2} depicts in the columns the three values of our study for the coefficients of determination of endogenous constructs.
Outliers are omitted in this graphic in order to expose changes in the average prediction performance more prominently.
As before, the gray horizontal lines represent the overall mean \gls{rmse} value of our study.
The results show that for higher values of the $R^2$, the \gls{rmse} values decrease. 
This holds for all prediction procedures and all \glspl{dgp} of our study.

Lastly, we assess to what extent the previous findings depend on the restriction on admissible estimation results.
For this purpose, we examine inadmissible Monte Carlo runs. 
The percentage of Monte Carlo runs that produce inadmissible results is presented in Table~\ref{tab:inadmissible}.
While the parameter estimation with \gls{plspm} yields no inadmissible results whatsoever, scenarios with dissimilar composites pose a challenge for both \gls{ml} and \gls{plsc} estimation.
Overall, \gls{plsc} produces the highest percentage of inadmissible results in all scenarios.
\begin{table}[htp!]
    \centering
    \begin{tabular}{lrrr}
    \toprule
     & \gls{ml} & \gls{plspm} & \gls{plsc} \\
     \midrule
     Common Factor (FFFF) & 0 & 0 & 15 \\
     Similar Composite (CCCC) & 0 & 0 & 11 \\
     Dissimilar Composite (CCCC) & 23 & 0& 25\\
     \bottomrule
    \end{tabular}
    \caption{Percentage of inadmissible results in our Monte Carlo study.}
    \label{tab:inadmissible}
\end{table}
To assess how excluding those inadmissible runs affects predictive performance, we also report the results using all Monte Carlo runs, including both admissible and inadmissible results in Appendix~\ref{app:allEstimates}.
Overall, the predictive performance and consequently our previous findings remain unchanged.
However, some of the inadmissible Monte Carlo runs exacerbate the issue of exceptionally large prediction errors.

\clearpage

\section{Discussion} \label{sec:discussion}

Prediction is central to many research areas, making it important to understand the capabilities and limitations of different procedures used to generate predictions.
Three such procedures for out-of-sample and operative prediction in \gls{sem} are MLpredict, PLSpredict, and PLScpredict.
Although each has been evaluated individually, their predictive performance has not yet been directly compared.
In this work, we thus conducted a Monte Carlo study in which we compared the three proposed procedures across a wide range of design conditions, including the type of constructs, and the coefficients of determination of endogenous constructs.
Overall, our results show that MLpredict, PLSpredict and PLScpredict exhibit similar predictive performance across all evaluation criteria and all considered design conditions. 
This finding suggests that none of the procedures is uniformly superior for generating operative predictions.
Despite the literature’s emphasis on \gls{pls} being causal-predictive \citep{Chin2020DemystifyingResearch}, our findings suggest that both MLpredict and PLS(c)predict perform similarly for this task.
An exception occurs in a small number of conditions, in which PLScpredict produced substantially larger prediction errors.
Because these extreme prediction errors are observed neither with MLpredict nor with PLSpredict, they might be attributable to the correction for attenuation, especially when it is inadequate.
Although this correction is beneficial for explanatory purposes, similar outliers in prediction errors have been reported for data with ordinal indicators \citep{Schamberger2023aMan}, even when constructs were specified correctly as common factors. 
This result further underlines that a procedure’s benefits for estimating model parameters need not translate directly into superior predictive performance.

The in-depth analysis of the individual design factors suggests that prediction quality is driven mainly by properties of indicators and constructs rather than by the choice of a procedure.
The configuration of the \glspl{dgp} was the most important factor for the predictive performance.
The comparison of \glspl{dgp} shows that when the FFFF and CCCC \glspl{dgp} were highly similar, all three procedures yielded lower \glspl{rmse} than when the CCCC configurations were dissimilar. 
Because all prediction procedures show a similar pattern regardless of construct misspecification, the differences in prediction error are more likely related to properties of the \glspl{dgp} than to construct misspecification itself.
In particular, the dissimilar composites in our study have, on average, lower correlations with their associated indicators, which may contribute to the worse predictive performance.
Moreover, greater $R^2$ values of $\eta_3$ and $\eta_4$ were associated with smaller prediction errors.
In contrast, the training sample size has little effect on the predictive performance. 
Presumably, all sample sizes in our study were sufficiently large, rendering additional observations incapable to further reduce out-of-sample prediction error.
This somewhat opposes the broader prediction literature, which claims larger sample requirements for prediction \citep{Shmueli2010ToPredict}.
Moreover, the correlation~$\rho$ of the exogenous constructs did not appear to affect the prediction performance, despite reflecting the redundancy in the information they provide through their associated indicators.
In our study design, $\rho$ has been used to derive the population path coefficients that maintain a given $R^2$ value.
Particularly, higher values for $\rho$ were accompanied by smaller path coefficients.
Presumably, this set-up contributed to $\rho$ not having a discernible effect on the predictive accuracy.
Furthermore, all procedures were not affected by whether the data was normally distributed or skewed.
While \gls{pls}-based procedures make no distributional assumptions \citep[\eg][]{Hair2022PartialExample}, MLpredict relies on normality of the indicators.
Against this background, its robustness against violations of the normality is noteworthy.
One possible explanation is that the method used to generate non-normal data can produce distributions that remain relatively close to Gaussian \citep{Foldnes2015HowApproach}.
These findings are likely to extend beyond the restriction on admissible results.
However, predictions based on inadmissible results were more likely to be exceptionally poor, reinforcing the practical advice to discard them.

Our findings are based on a simulation study, which is naturally limited by its design and thus open avenues for future research.
First, the simulation design was restricted to specific structural equation models.
Second, we restricted PLScpredict and MLpredict to common factor model specifications, even though both procedures are theoretically capable of handling composites as well \citep[\eg][for \gls{ml} estimation]{Schamberger2025Factor-Framework}.
Third, we examined operative prediction only, leaving the performance for alternative types of prediction, \eg\ latent prediction for future research. 
While both PLS-based procedures are, in principle, easily adaptable to other prediction types, MLpredict has been developed specifically for operative prediction. 
However, neither PLS-based prediction nor MLpredict is currently available in the widely used statistical software \textsf{R} for any type of out-of-sample prediction other than operative.
Future work should therefore generalize MLpredict to broader prediction settings, and provide open-source, ready-to-use implementations for both MLpredict and PLS(c)predict.

\section{Conclusion}  \label{sec:conclusion}
We compared the prediction performance of three procedures to compute out-of-sample predictions in \gls{sem} by means of a comprehensive Monte Carlo simulation and found none of the considered methods superior for operative prediction.
Overall, our results emphasize that, for practitioners, investing effort in identifying indicators that are strongly associated with their constructs, and specifying a model with strong structural relations has a great impact on the predictive performance, whereas the effects of construct misspecification or the choice for a particular prediction procedure seem of little importance.
While this finding might appear surprising, given the severe effects of construct misspecification on explanatory performance \citep[\eg][]{Bauer2025MisspecificationsCompositesArxiv, Cho2022ARepresentations, Aguirre-Urreta2014, Sarstedt2016}, it aligns with the idea that prediction often tolerates certain forms of model misspecification as long as the main predictive signal can be captured, \ie\ trading some bias for reduced prediction variance \citep{Hastie2009TheLearning, Shmueli2010ToPredict}. 
The larger prediction errors occasionally observed for PLScpredict illustrate the tension between consistent parameter estimation (for explanation) and minimized prediction error, stressing the broader distinction between explanatory and predictive modeling in \gls{sem}. 
In this way, our study guides applied researchers in balancing these two goals in their work.
Moreover, it complements the growing literature on predictive \gls{sem} \citep[\eg][]{Tan2025DeterminantsStudy, Stefko2022PsychologicalNegligible, Taufiq-Hail2023InvestigatingPLSpredict}, which has so far focused mainly on PLSpredict \citep[\eg][]{Hair2021Man}, by demonstrating that the claimed predictive advantages of \gls{pls}-based methods do not translate into clear performance differences.

The findings of this study have several implications for empirical researchers who rely on \gls{sem} for prediction. 
Given the largely similar predictive performance of PLSpredict, PLScpredict, and MLpredict, fine-grained methodological distinctions between these procedures may be less consequential when the primary goal is accurate out-of-sample prediction rather than explanation. 
In practice, this suggests that researchers should focus on accurate model specification, measurement quality, and transparency in reporting predictive diagnostics, provided that any chosen procedure is applied and interpreted correctly.  
Overall, our results underline that methodological choices and construct misspecification have fundamentally different consequences for prediction than for explanation and thus emphasize the conceptual distinction between explanatory and predictive modeling goals in \gls{sem}.
\newpage
\section*{Declarations}
\subsection*{Author contribution statement}
\begin{table}[htp!]
    \centering
    \renewcommand{\arraystretch}{1.5}
    \begin{tabular}{lp{8cm}}
         \toprule
         Jonas Bauer\orcidlink{0000-0003-1759-7889}& Conceptualization, Methodology, Investigation, Formal analysis, Software, Visualization, Writing – Original Draft\\
         Florian Schuberth\orcidlink{0000-0002-2110-9086} &  Conceptualization, Writing – Review \& Editing \\
         Christiane Fuchs\orcidlink{0000-0003-3565-8315}& Supervision, Writing -- Review \& Editing \\
         Tamara Schamberger\orcidlink{0000-0002-7845-784X} & Supervision, Conceptualization, Methodology, Investigation, Writing – Original Draft, Writing – Review \& Editing
         \\ \bottomrule
    \end{tabular}
    \label{tab:placeholder}
\end{table}

\subsection*{Data availability statement}
All \textsf{R} source files, the estimation results and Supplementary Material are available at \href{https://doi.org/10.5281/zenodo.23506099}{https://doi.org/10.5281/zenodo.23506099}.
\newpage
\printbibliography
\newpage
\appendix

\section{Deriving path coefficients from coefficients of determination}
\label{appendix:path-coefs}

The path coefficients used in the Monte Carlo simulation are derived based on the design factors $R^2$ and $\rho$. 
More precisely, we derive standardized path coefficients under the restriction that both endogenous constructs $\eta_{3}$ and $\eta_{4}$ should have unit variance, and such that the resulting coefficients of determination match the design factor values.

Suppose the two exogenous constructs $\eta_1$ and $\eta_2$ have unit variance and correlation $\text{Cor}(\eta_{1}, \eta_{2}) = \rho$. The first endogenous construct $\eta_{3}$ follows the structural equation
\[
\eta_{3} = \beta_{1} \eta_{1} + \beta_{2} \eta_{2} + \zeta_{3},
\]
where the coefficients of the paths from both exogenous constructs to $\eta_{3}$ are constrained to be equal, \ie\ $\beta_{1} =\beta_{2}$.
The disturbance term $\zeta_{3}$ is assumed to be uncorrelated with all constructs, and its variance is specified such that $\text{Var}(\eta_{3}) = 1$.

Under these assumptions, the variance~$p_{\eta_3}$ explained in $\eta_{3}$ is
\[
p_{\eta_3} = \text{Var}\bigl( \beta_{1} \eta_{1} + \beta_{2} \eta_{2} \bigr) =
\beta_{1}^{2} \bigl( \text{Var}(\eta_{1}) + \text{Var}(\eta_{2}) + 2  \text{Cov}(\eta_{1}, \eta_{2}) \bigr).
\]
Using $\text{Var}(\eta_{1}) = \text{Var}(\eta_{2}) = 1$ and $\text{Cov}(\eta_{1}, \eta_{2}) = \rho$, this simplifies to
\[
p_{\eta_3} = 2(1+\rho) \beta_{1}^{2}.
\]
The target $R^{2}$ is then equated to the explained variance:
\[
R^{2} = p_{\eta_3} = 2(1+\rho) \beta_{1}^{2},
\]
yielding
\begin{align}
    \beta_{1} = \sqrt{\frac{R^{2}}{2(1+\rho)}}
    . \label{eq:pathCoefDet}
\end{align}
The residual variance of $\eta_{3}$ is given by
\[
\text{Var}(\zeta_{3}) = 1 - p_{\eta_3} = 1 - R^{2}.
\]

The second endogenous construct is specified as
$$
\eta_4
=
\beta_{4}\eta_3
+
\beta_{3}\eta_2
+
\zeta_4,
$$
where $\zeta_4$ is uncorrelated with $\eta_2$ and $\eta_3$. Given
$\operatorname{Var}(\eta_2)=\operatorname{Var}(\eta_3)=1$, the variance~$p_{\eta_4}$ explained in $\eta_4$ is
$$
p_{\eta_4} = \operatorname{Var}(\beta_{4}\eta_3+\beta_{3}\eta_2)
= \beta_{4}^{2}
+ \beta_{3}^{2}
+ 2\beta_{4}\beta_{3}\underbrace{\operatorname{Cov}(\eta_3,\eta_2)}_{=\star}.
$$
The covariance in $\star$ is given as
$$
\operatorname{Cov}(\eta_3,\eta_2)
=
\beta_{1}\rho + \beta_{2} = \beta_{1}(1+\rho).
$$
The target $R^2$ is then imposed by solving
$$
\beta_{3}^{2}
+
\beta_{3} 2\beta_{4}\beta_{1}(1+\rho)
+
\beta_{4}^{2}
=
R^2.
$$
We set $\beta_{4}=s\sqrt{R^2}$ for a particular $s$ and select the root for $\beta_{3}$ with the smaller absolute value. 
For the default value $s=\frac{1}{2}$ we get
$$
\beta_{4} =
\frac{1}{2}\sqrt{R^2},
$$
and
$$
\operatorname{Cov}(\eta_3,\eta_2) =(1+\rho)  \sqrt{\frac{R^{2}}{2(1+\rho)}} =
\sqrt{\frac{R^2(1+\rho)}{2}}.
$$
Substituting these into the variance of the disturbance term~$\zeta_4$ of $\eta_4$ gives
\begin{align*}
\operatorname{Var}(\zeta_4)
&=
1-\operatorname{Var}(\beta_{4}\eta_3+\beta_3\eta_2)\\
&=
1-\left[
\beta_{4}^2+\beta_{3}^2
+2\beta_{4}\beta_{3}
  \operatorname{Cov}(\eta_3,\eta_2)
\right]\\
&=
1-\left[
\frac{R^2}{4}
+\beta_{3}^2
+2\left(\frac{\sqrt{R^2}}{2}\right)
  \beta_{3}
  \sqrt{\frac{R^2(1+\rho)}{2}}
\right]\\
&=
1-\left[
\frac{R^2}{4}
+\beta_{3}^2
+\beta_{3}R^2
  \sqrt{\frac{1+\rho}{2}}
\right].
\end{align*}
We choose $\beta_{3}$ such that the quantity in brackets equals the target $R^2$ by solving
$$
\beta_{3}^2
+
\beta_{3}R^2\sqrt{\frac{1+\rho}{2}}
-\frac{3R^2}{4}
=0.
$$

In \textsf{R}, the calculations are implemented in the helper function \texttt{calcPathCoefs}.

\section{Population parameters of FFFF \glspl{dgp} and similar CCCC \glspl{dgp}} \label{appendix:icm}

For every construct~$\eta_j$, its type determines the variance-covariance matrix of the associated indicators~$\bm x_j$ and the covariances between the construct and the indicators.
All constructs in our simulation have unit variance.
If $\eta_j$ is a common factor, the variance-covariance matrices of its associated indicators $\bm x_j$ and that of their measurement error terms~$\bm \varepsilon_j$ are given by
$$
\text{Var}(\bm x_j) = \bm\lambda_j\bm\lambda_j^\top + \bm \Theta_j \quad , \quad \text{Var}(\bm \varepsilon_j) = \bm \Theta_j = \text{diag}(\theta_{j,1}, \theta_{j,2}, \theta_{j,3}).
$$

\paragraph{Construction of common factors}

The covariances between a common factor and the associated indicators are parameterized by the factor loadings $\bm \lambda_j =(\lambda_{j,1}, \lambda_{j,2}, \lambda_{j,3})$.
We fix these (unstandardized) loadings to $\bm \lambda_j = (0.9, 0.8, 0.7)$.
All measurement errors in our study are uncorrelated and have variances of 0.3, yielding
$$
\text{Var}(\bm \varepsilon_j):=\bm \Theta_j = \begin{pmatrix}
    0.3 & 0 & 0 \\
    0 & 0.3 & 0 \\
    0 & 0 & 0.3
\end{pmatrix}.
$$

\paragraph{Construction of composites}
A composite~$\eta_j$ with mean zero (which is always the case in this work) would imply covariances between construct and indicators of $\bm \Sigma_j \bm w_j$, where all entries of the intra-block variance-covariance matrix~$\bm \Sigma_j$ of the indicators are model parameters.
The specification of this matrix to match properties from the common factor case can be achieved as follows:
\begin{align}
    \bm\Sigma_j = \bm\lambda_j\bm\lambda_j^\top + \bm \Theta_j. \label{app:compSigma}
\end{align}
Correlations between composite and indicators that are similar to the FFFF \glspl{dgp} can be achieved by setting
\begin{align}
    \tilde{\bm w_j} &= {\left( \bm\lambda_j\bm\lambda_j^\top + \bm \Theta_j\right)}^{-1} \bm \lambda_j.  \\
    {\bm w_j} &= \frac{\tilde{\bm w_j}}{\sqrt{\tilde{\bm w_j}^{\top}\bm{\Sigma_j}\tilde{\bm w_j}}}, \label{app:compW}
\end{align}
where $\tilde{\bm w_j}$ are the unnormalized weights.
\clearpage

\section{Population parameters with 4 indicators}

With $K=4$ indicators associated with every construct, the factor loadings are specified as
$$
\bm \lambda_j = \begin{pmatrix}
    0.9 \\ 0.8 \\ 0.7 \\ 0.6
\end{pmatrix}
$$
and the measurement error variances as
$$
\bm \Theta_j = \begin{pmatrix}
    0.3 & 0 & 0 & 0\\
    0 & 0.3 & 0 & 0\\
    0 & 0 & 0.3 & 0 \\
    0 & 0 & 0 & 0.3
\end{pmatrix}.
$$
For composites, the weights and intra-block variance-covariance matrices are given as

\begin{align}
    \bm w_j \approx &\begin{pmatrix}
        0.37 \\ 0.33 \\ 0.29 \\ 0.25
    \end{pmatrix} \ , \bm \Sigma_j = \begin{pmatrix}
        1.11 & 0.72 & 0.63 & 0.54 \\
    0.72& 0.94& 0.56 & 0.48\\
    0.63 &0.56 &0.79& 0.42 \\
    0.54 &0.48& 0.42& 0.66
    \end{pmatrix} \tag{similar} \\
    \bm w_j =& \begin{pmatrix}
        0.15 \\0.30 \\0.45 \\0.60
    \end{pmatrix} \ , \bm \Sigma_j \approx \begin{pmatrix}
        0.98 &0.19& 0.22 &0.25\\
    0.19 &0.91& 0.28& 0.30 \\
    0.22 &0.28& 0.84& 0.32 \\
    0.25 &0.30 &0.32 &0.77
    \end{pmatrix}  \tag{dissimilar} 
\end{align}

\clearpage

\section{Additional visualizations of the simulation results} \label{app:sim_res}
\begin{figure}[ht]
    \centering
    \includegraphics[width=\linewidth]{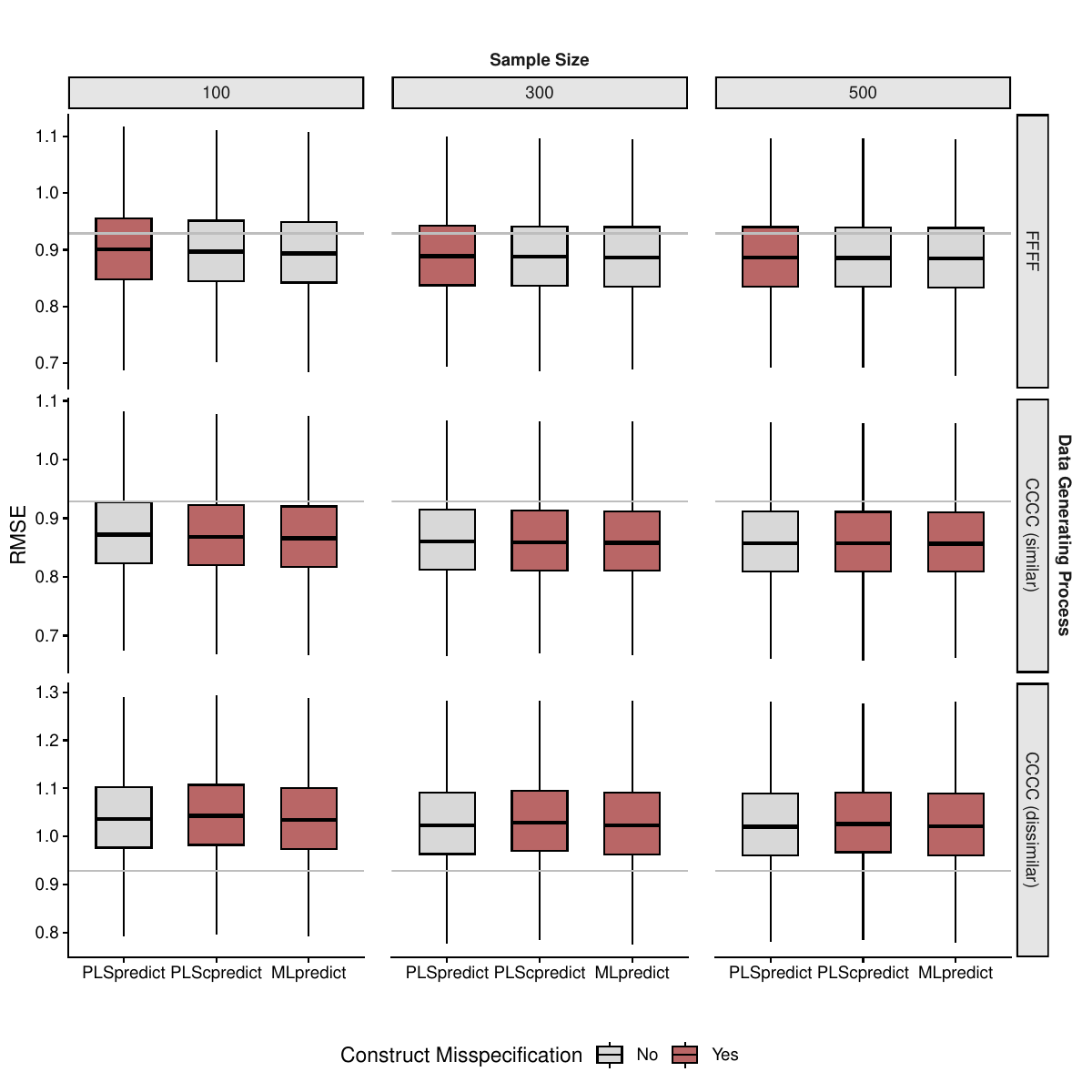}
    \caption{\gls{rmse} of operative predictions for all indicators associated with endogenous constructs with respect to the different training sample sizes of our study. Red boxes indicate conditions involving construct misspecification. The gray horizontal line represents the overall mean \gls{rmse} value of our study.}
    \label{app:fig:fig4SampleSize}
\end{figure}
\begin{figure}[ht]
    \centering
    \includegraphics[width=\linewidth]{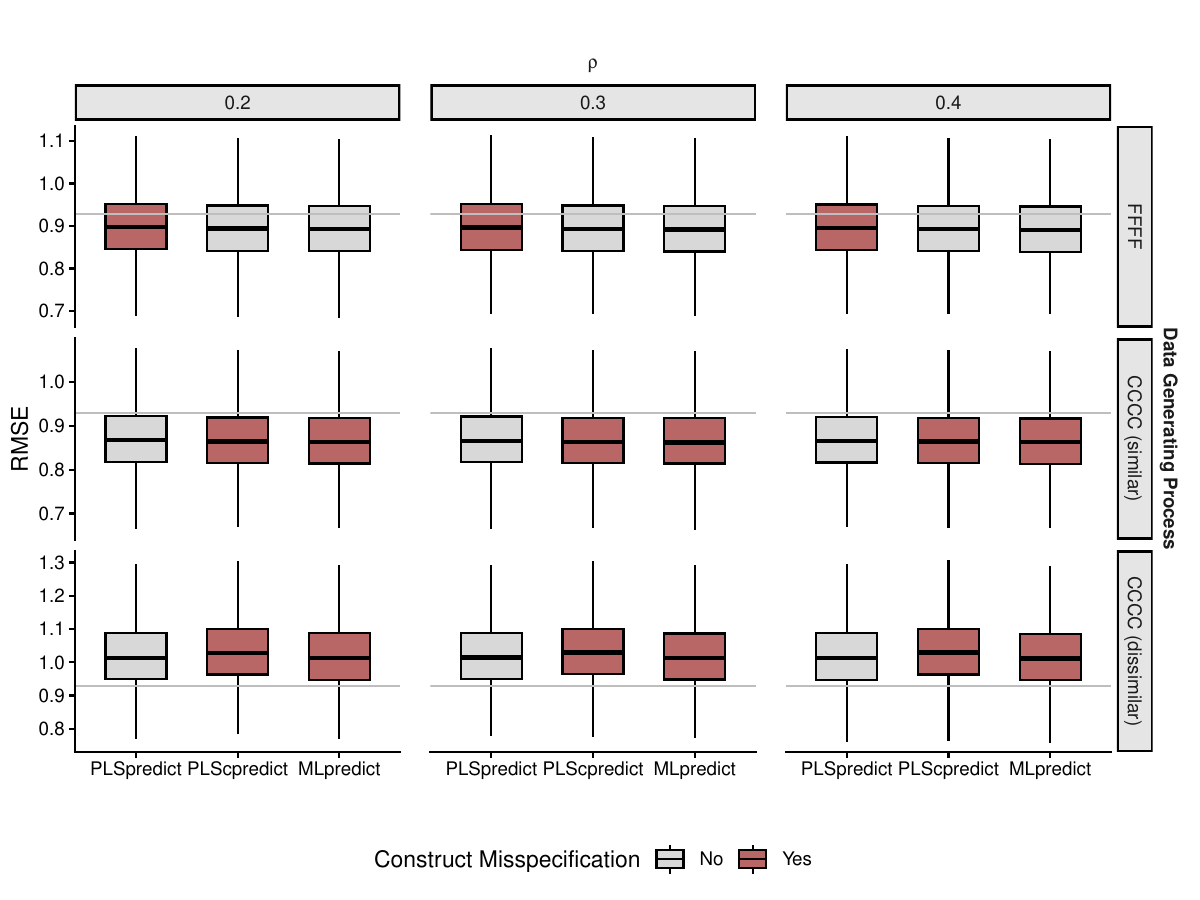}
    \caption{\gls{rmse} of operative predictions for all indicators associated with endogenous constructs with respect to the correlation $\rho$ between the two exogenous constructs (in the columns) of our study. Red boxes indicate conditions involving construct misspecification. The gray horizontal line represents the overall mean \gls{rmse} value of our study.}
    \label{app:fig:fig4CorEx}
\end{figure}
\begin{figure}[ht]
    \centering
    \includegraphics[width=0.75\linewidth]{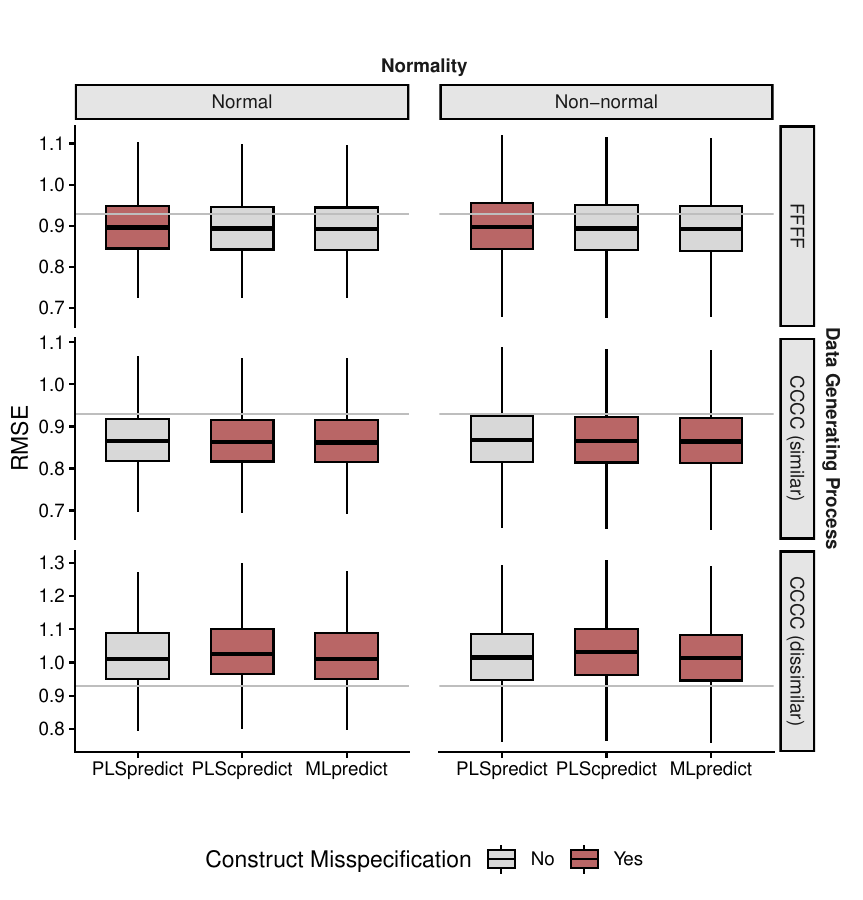}
    \caption{\gls{rmse} of operative predictions for all indicators associated with endogenous constructs with respect to the normality of the indicators (in the columns). Red boxes indicate conditions involving construct misspecification. The gray horizontal line represents the overall mean \gls{rmse} value of our study.}
    \label{app:fig:fig4Normality}
\end{figure}

\clearpage

\section{Visualizations including inadmissible results} \label{app:allEstimates}

\begin{figure}[ht!]
    \centering
    \includegraphics[width=0.9\linewidth]{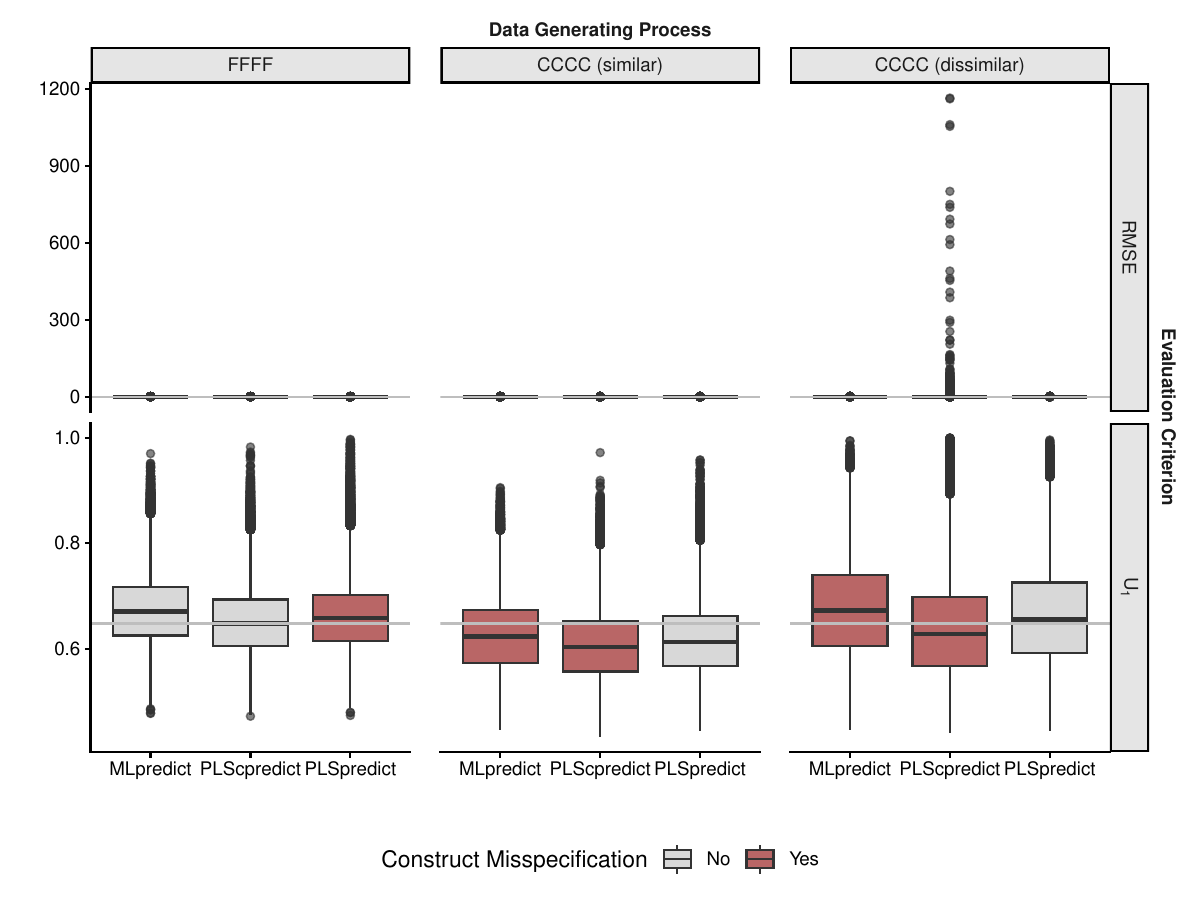}
    \caption{\gls{rmse}, and $U_1$ of MLpredict, PLSpredict, and PLScpredict of all predicted indicators and all Monte Carlo conditions with data from common factor models (FFFF, left column), similar composite models (CCCC; center) and dissimilar composite models (CCCC; right). The boxplots include outliers. Red boxes indicate conditions with construct misspecification.}
    \label{fig:app:operative_allEstimateswOutliers}
\end{figure}

\begin{figure}[ht!]
    \centering
    \includegraphics[width=0.9\linewidth]{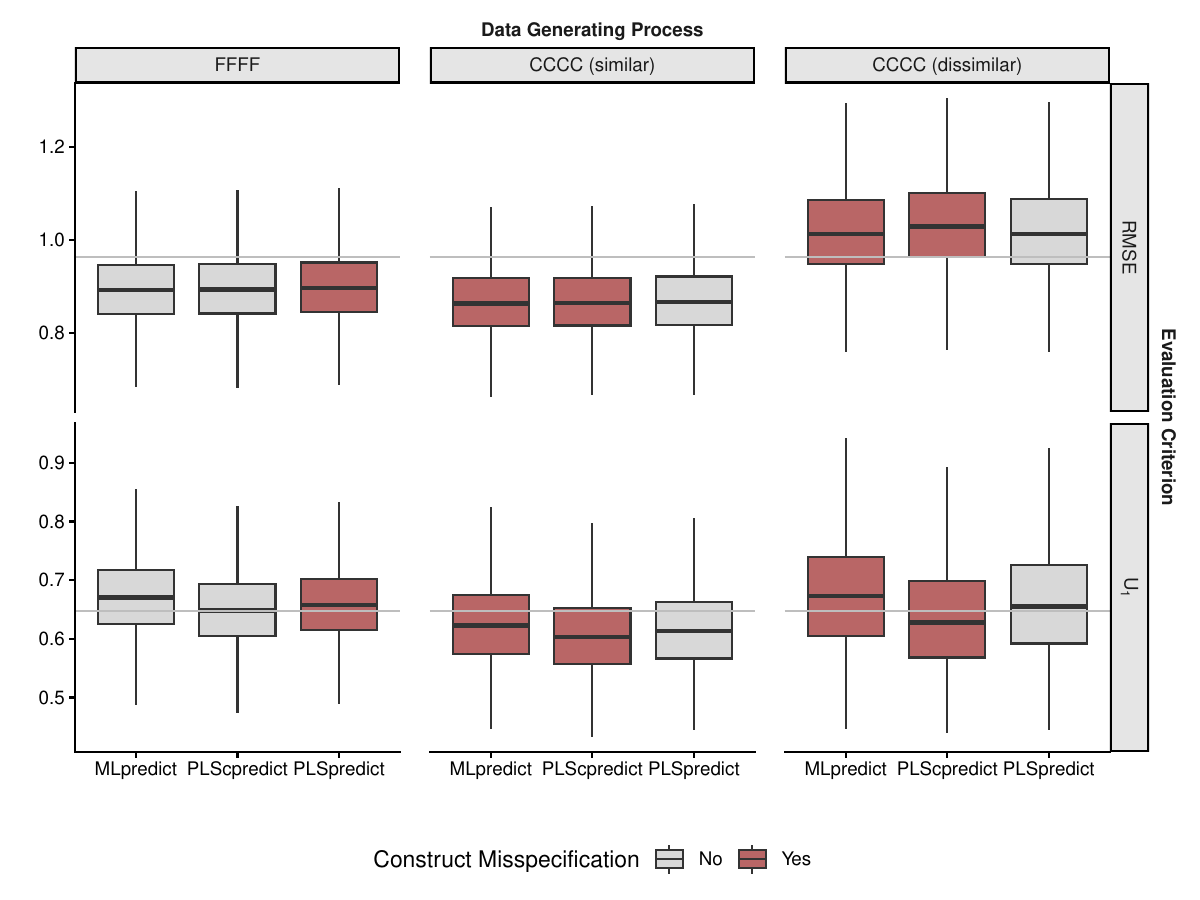}
    \caption{\gls{rmse}, and $U_1$ of MLpredict, PLSpredict, and PLScpredict of all predicted indicators and all Monte Carlo conditions with data from common factor models (FFFF, left column), similar composite models (CCCC; center) and dissimilar composite models (CCCC; right). . The boxplots exclude outliers. Red boxes indicate conditions with construct misspecification.}
    \label{fig:app:operative_allEstimatesWithoutOutliers}
\end{figure}

\begin{figure}[htp!]
    \centering
    \includegraphics[width=\linewidth]{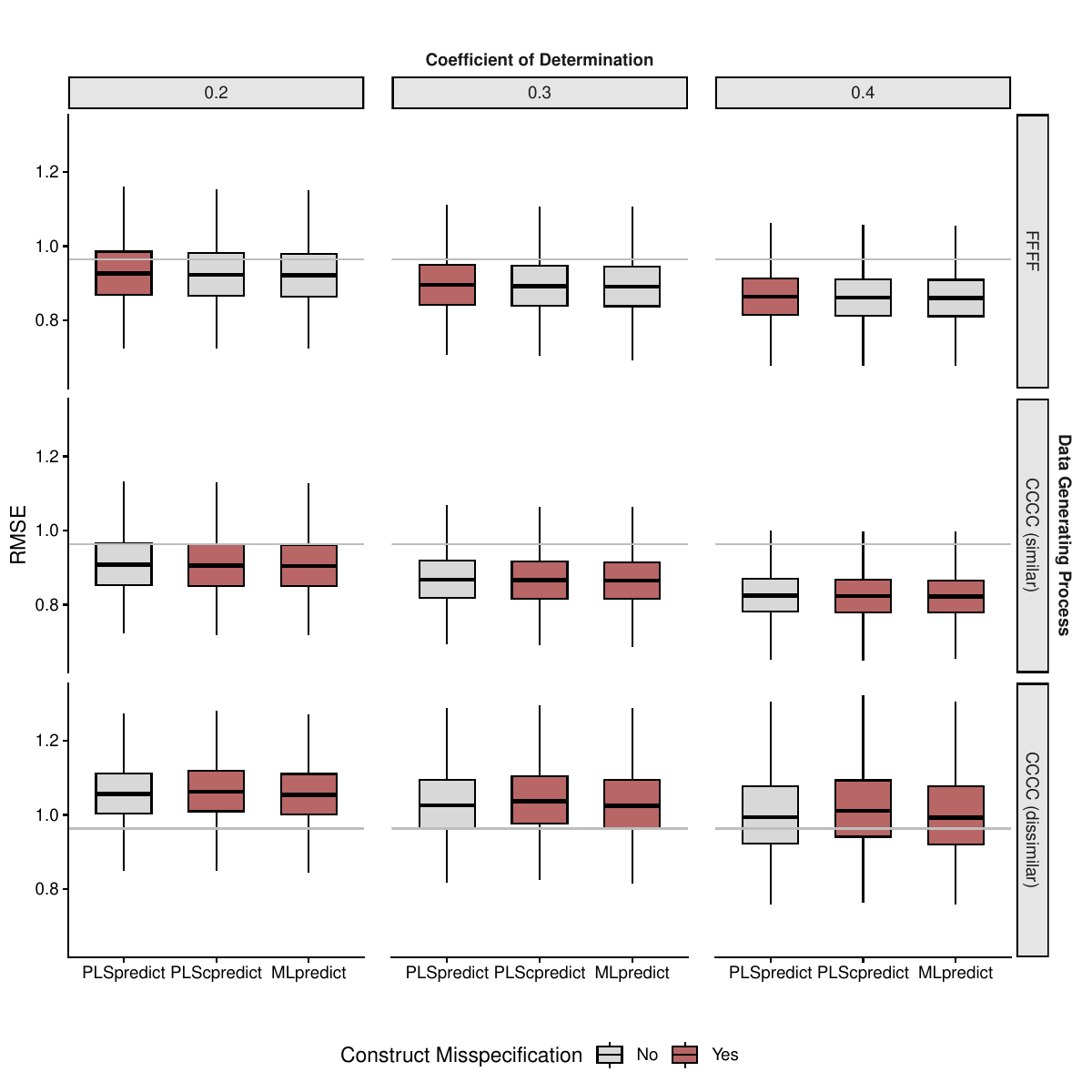}
    \caption{\gls{rmse} of operative predictions for all indicators associated with endogenous constructs with respect to the $R^2$ values (in the columns) of our study. Red boxes indicate conditions involving construct misspecification.}
    \label{fig:app:determination}
\end{figure}

\begin{figure}[htp!]
    \centering
    \includegraphics[width=\linewidth]{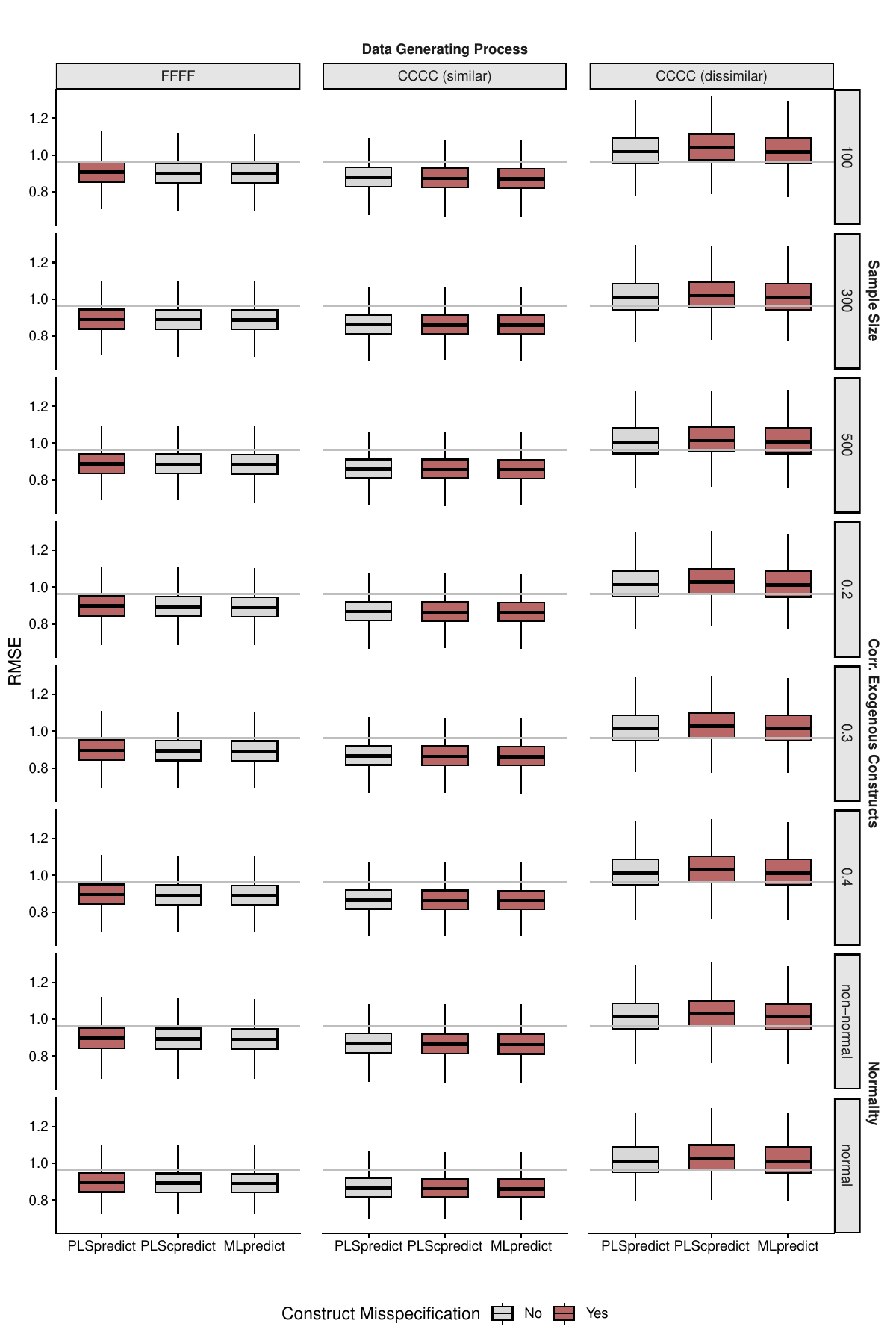}
    \caption{\gls{rmse} of operative predictions for all indicators associated with endogenous constructs with respect to several design factors of our study (in the rows). Red boxes indicate conditions involving construct misspecification.}
    \label{fig:app:indRel_loadings}
\end{figure}

\end{document}